\documentclass[aps,prl,twocolumn]{revtex4-1}
\usepackage{graphics}  
\usepackage{epsfig}
\usepackage[usenames]{color}
\usepackage{verbatim}
\usepackage{amsmath}

\IfFileExists{srcltx.sty}{\usepackage[active]{srcltx}}

\begin{document}

\title{Solitosynthesis induced phase transitions}


\author{Lauren Pearce}
\affiliation{Department of Physics and Astronomy, University of California, Los Angeles, CA 90095-1547, USA}

\begin{abstract}
We consider a phase transition induced by the growth of Q-balls in a false vacuum.  Such a transition could occur in the early universe in the case of broken supersymmetry with a metastable false vacuum.  Small Q-balls with a negative potential energy can grow in a false vacuum by accretion of global charge until they reach critical size, expand, and cause a phase transition.  We consider the growth of Q-balls from small to large, using the  Bethe-Salpeter equation to describe small charge solitons and connecting to the growth of larger solitons for which the semiclassical approximation is reliable.  We thus test the scenario in a simplified example inspired by supersymmetric extensions of the standard model. 
\end{abstract}

\maketitle

\section{Introduction}

Q-balls~\cite{Coleman} are non-topological solitons~\cite{NTS,NTS1} that are stable because they carry a conserved global charge.   They arise in a number of models, and, in particular, in supersymmetric extensions of the Standard Model, where they carry baryon  and/or lepton number~\cite{Kusenko MSSM Qballs}.  Stable supersymmetric Q-balls can form in the early universe from the fragmentation of Affleck-Dine condensate~\cite{Affleck-Dine,KS} or in other processes~\cite{solitogenesis}, and they can play the role of cosmological dark matter~\cite{KS,DM}. 
Furthermore, it has been suggested that Q-balls can facilitate phase transitions even when the tunneling rate is too small for the phase transition to occur otherwise; the Q-balls accumulate charge until they reach a critical charge, at which point they expand, causing a phase transition~\cite{Kusenko Solitosynthesis}.  Such a phase transition could have interesting cosmological implications.

While the possibility of such a phase transition has been explored in the literature~\cite{Postma, Metaxas1}, a complete model of it has not yet been demonstrated.  This is due to difficulties with the quantum nature of small charge Q-balls and also with the properties of Q-balls in the false vacuum.  This paper will demonstrate, from beginning to end, a model in which a phase transition is induced by solitosynthesis of Q-balls.

This paper is organized as follows: first we specify the potential that gives rise to our Q-balls and show that it has the requisite properties; then we consider the properties of the non-topological solitons in the false vacuum.  There are primarily four regimes to consider.  For large charges, the thin wall semiclassical approximation is valid, while for smaller charges, the thick wall semiclassical approximation is valid.  For intermediate charges, we interpolate between these two regimes.  For extremely small charges, quantum effects are important and the semiclassical approximation is invalid; instead, we apply the Bethe--Salpeter equation.

After we have described the radii and energies of the Q-balls, we proceed to consider the properties of the phase transition; in particular, the critical charge and the critical radius.  Then we consider solitosynthesis, the process by which Q-balls grow by accreting of charge.  We find the temperature at which such growth begins, and then we calculate the rate of growth in each regime.  We demonstrate that the growth is not hindered by charge depletion and freeze out, which could end solitosynthesis before critically sized Q-balls form.  Finally, we discuss explicit numerical examples to show that such a phase transition is a theoretical possibility.

In all this, we use a simplified toy model inspired by the Minimal Supersymmetric Model (MSSM).  In the last section, we discuss the application of this analysis to the MSSM, and in particular we consider phase transitions of cosmological interest.

\section{The potential}

\begin{figure}
\includegraphics[scale=.7]{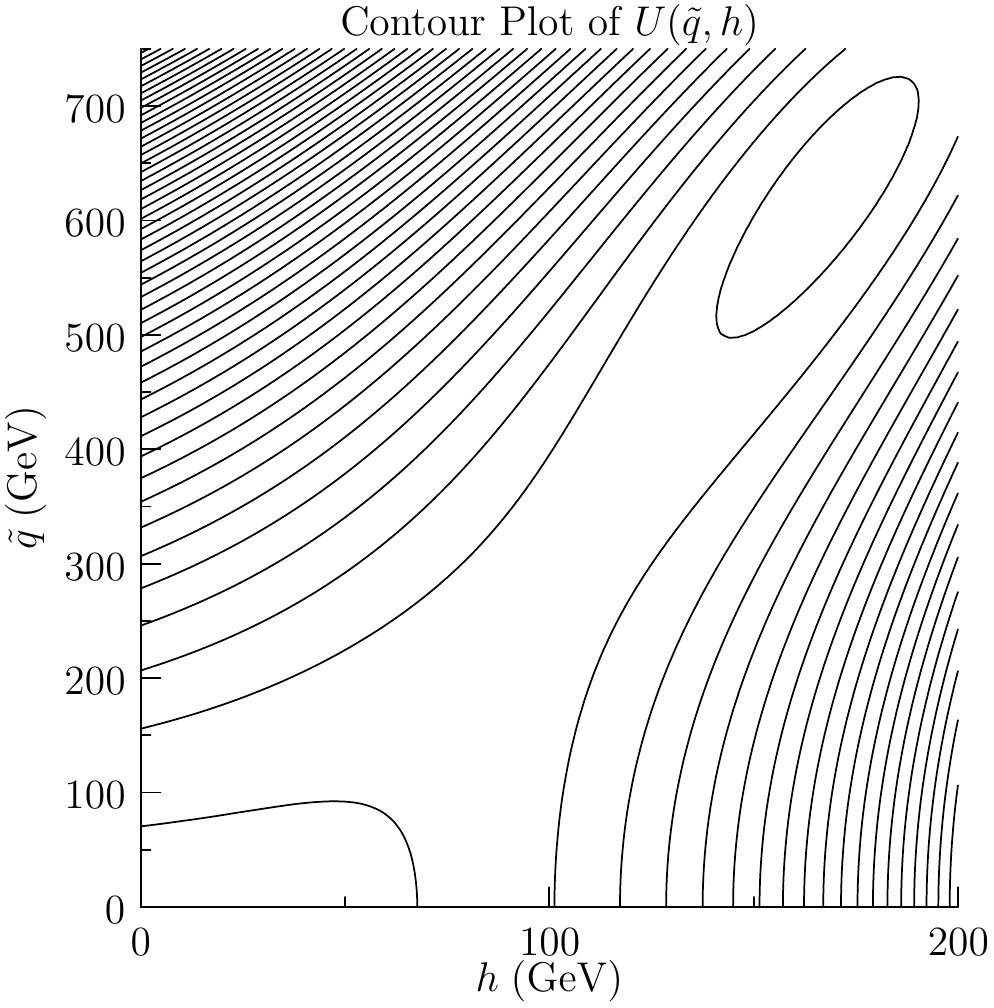}
\caption{A contour plot of the potential $U(\tilde{q},h)$.  There is a local minimum at the origin and a global minimum at $\left< h \right> = 169$~GeV and $\left< \tilde{q} \right> = 631$~GeV.} \label{fig:UContour}
\end{figure}

For this paper, we will use an MSSM-inspired potential~\cite{Kusenko Solitosynthesis,Postma}:
\begin{align}
U = \dfrac{m_q^2}{2} \tilde{q}^2 + \dfrac{m_h^2}{2} h^2 - A_0 h \tilde{q}^2 +  \dfrac{\lambda_1}{4} \tilde{q}^2 h^2 + \dfrac{\lambda_2}{4} \tilde{q}^4 + \dfrac{\lambda_3}{4} h^4,
\label{eq:MSSMPotential}
\end{align}
in which $\tilde{q}$ is a squark field and $h$ is the lightest Higgs boson.  For simplicity, we use real fields.  In general, renormalization effects, including the effects of Q-balls if they exist, can be significant~\cite{Renormalization}; we take the couplings to be the renormalized couplings.  We assume that any other particles which carry baryon number are heavier than the squark, to ensure the squark's stability.  The origin is always a local minimum of this potential; however, for particular values of the coupling constants a different global minimum exists.  For example, if $m_q = 200$~GeV, $m_h = 10$~GeV, $A_0 = 240$~GeV, $\lambda_1=\lambda_2=.1$, and $\lambda_3=19$, the  origin is a local minimum and there are global minima at $\left< h \right> = 169$~GeV and $\left< \tilde{q} \right> = \pm631$~GeV.  In Fig. \ref{fig:UContour}, we show a contour plot of this potential. 

\begin{figure}
\includegraphics[scale=.7]{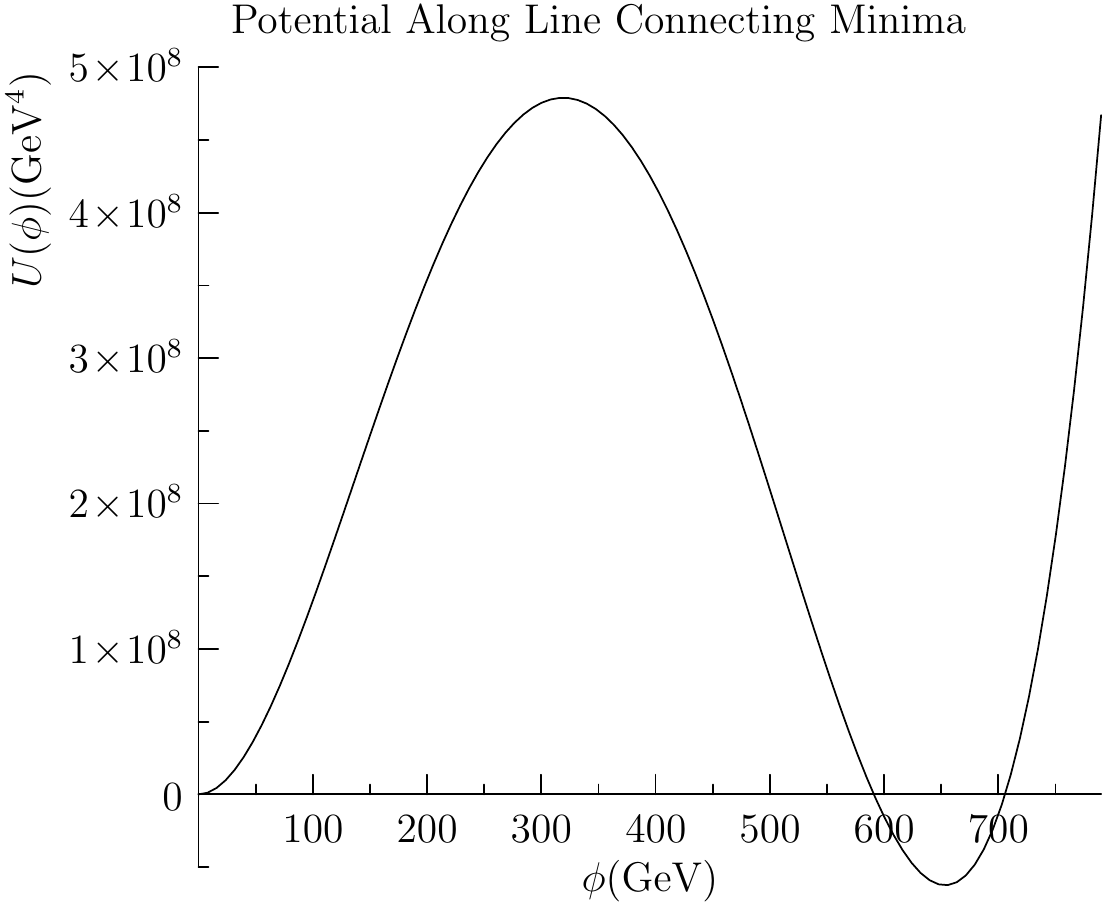}
\caption{This is the potential along the line connecting the false vacuum to the true vacuum.  We see a large barrier, which suppresses phase transitions driven by thermal fluctuations.} \label{fig:ULineMin}
\end{figure}

Phase transitions involving multiple fields are difficult to solve exactly; a reasonable approximation is that they occur along the line connecting the two minima.  This is a valid approximation when the potential does not have an unusually shaped barrier between the minima; we see no sign of an unusual barrier in the contour plot.  The potential along this line may be found by substituting $\tilde{q} = \phi \sin(\theta) $ and $\tilde{h} = \phi \cos(\theta)$ with $\theta = 1.309 $:
\begin{equation}
U(\phi) = 18700 \; \mathrm{GeV}^2 \; \phi^2 - 58.0 \; \mathrm{GeV} \; \phi^3 + .0447 \phi^4 \end{equation}
which is shown in Fig. \ref{fig:ULineMin}.  We see that there is a large barrier between the minima; this suppresses phase transitions induced by thermal fluctuations.

Next we must demonstrate the existence of Q-balls in the false vacuum (located at the origin).  The conserved charge carried by our Q-balls will be baryon number; this is conserved in the false vacuum because squarks do not have a vacuum expectation value.  The condition for the existence of Q-balls involving multiple fields is that $2U \slash \sum_k Q_k \phi_k^2$ is minimized at a nonzero value of the fields, where $Q_k$ is the charge of of the field $\phi_k$~\cite{Coleman,Kusenko MSSM Qballs}.  Because the baryon number carried by the squark field is $1 \slash 3$, we consider the minimum of $6 U \slash \tilde{q}^2$.  Due to the $\tilde{q}^2 h$ term, the origin will not be even a local minimum of $U\slash \tilde{q}^2$; in fact, the global minimum is located at $\tilde{q}_0 = 624$~GeV and $h_0=168$~GeV, and in which the potential is $-6.18\cdot 10^7 \; \mathrm{GeV}^4$.

Therefore, Q-balls carrying baryon number exist in this vacuum, and furthermore, the potential is negative within their interior.  This is necessary for a Q-ball induced phase transition to occur; such a phase transition converts the fields everywhere to their values inside of the Q-ball, $\tilde{q}_0$ and $h_0$.  Since this is not quite the true minimum of the potential, the system will then slide classically into the true minimum.

Again we consider the potential along the line connecting the initial vacuum and the final state, which, with $\tilde{q} = \phi \sin(\theta) $, $h = \phi \cos(\theta)$, and $\theta = 1.308 $, is:
\begin{equation}
U(\phi) = 18700 \; \mathrm{GeV}^2 \; \phi^2 - 58.2 \; \mathrm{GeV} \; \phi^3 + .0450 \phi^4.
\label{eq:Potential}
\end{equation}
We can also describe the Q-balls in terms of the $\phi$ field.  The field inside such a Q-ball is the value of $\phi$ that minimizes $U(\phi) \slash \phi^2$, which is $\phi_0 = 646$~GeV, and as expected, this is $\sqrt{ \tilde{q}_0^2 + h_0^2} $.  Also, as expected, $U(\phi_0) = - 6.18 \cdot 10^7 \; \mathrm{GeV}^4$.  Thus, at least for large charge, we can consider our Q-balls as being bound states of squarks exchanging Higgs bosons, or as coherent oscillations of $\phi$ quantum.  For small charge states, we should remember that the charge is really carried by squarks that exchange Higgs bosons.

Finally, because squarks carry charge $1 \slash 3$, the charge present in a field configuration is:
\begin{align}
Q &=\dfrac{\omega}{3} \int d^3x \, \tilde{q}^{2} =\dfrac{\sin^2(1.308)}{3} \omega \int d^3x \, \phi^2,
\end{align}
where $\omega$ describes the oscillatory time dependence of the fields.  For later convenience, we measure the charge in units such that $Q = \sin^2(1.308) Q^\prime \slash 3$; then:
\begin{equation}
Q^\prime =  \omega \int d^3x \, \phi^2.
\end{equation}
Physically, $Q^\prime$ is the charge in a single $\phi$ quanta.  In terms of units of charge $Q^\prime$, a single squark carries charge 3.107; this will be even closer to 3 the nearer the global minimum is to the $\tilde{q}$ axis.  Speaking loosely, since $\theta \approx \pi \slash 2$, the field $\phi$ is ``almost" the squark field, and the charge $Q^\prime$ is approximately the number of squarks present.

Generically, then, we consider a potential of the form
\begin{equation}
U(\phi) = \dfrac{m_0^2}{2} \phi^2 - A \phi^3 + \dfrac{\lambda}{4} \phi^4,
\end{equation}
in which the field $\phi$ is made of squarks and higgs bosons, and furthermore, the charge of the $\phi$ field is approximately the charge of the squark.  The general form of the relation between the charges is:
\begin{equation}
Q^\prime = \dfrac{3Q}{\sin^2(\theta)}
\label{eq:Charges}
\end{equation}

\subsection{Corrections to the Potential}

At the scale of color confinement, we expect the squarks to arrange into color singlets of the form $ \epsilon_{abc} \epsilon_{\alpha \beta} \tilde{Q}^\alpha_a \tilde{Q}^\beta_b \tilde{Q}_c $, where Greek letters denote $ SU(2) $ indices and Latin letters denote color indices ~\cite{gauge_structure_of_Q-balls}.  To avoid this complication, we will ensure that the relevant temperatures are above the scale of the QCD phase transition; furthermore, the scalar binding interaction mediated by Higgs bosons is much stronger than the strong interaction.  

In general, finite temperature corrections to the potential, $\lambda^2 T^2 \phi^2$ and $\lambda T \phi^3$, may be important; however, when the relevant temperatures are significantly smaller than $m_0$ and $A$, these finite temperature effects can be neglected.  This is the case in the examples considered. 

\section{Properties of Q-balls in the False Vacuum}

The exact and general equation for the energy of a Q-ball of arbitrary charge has three terms:
\begin{equation}
E(Q^\prime) = \int d^3x \left( \dfrac{1}{2} |\dot{\phi}|^2 + \dfrac{1}{2} |\nabla \phi |^2 +U(\phi) \right). \label{eq:EnergyBasic}
\end{equation}
The field oscillates in time as $e^{\imath \omega t} \bar{\phi}(x)$ where the frequency is related to the charge according to:
\begin{equation}
Q^\prime = \dfrac{1}{2\imath} \int d^3x \, \phi^* \overleftrightarrow{\partial_t} \phi = \omega \int \phi^2 \, d^3 x .
\end{equation}
After some manipulation, one can write~\cite{Kusenko Small Qballs}:
\begin{align}
E &= \int d^3x \left( \dfrac{1}{2} | \nabla \bar{\phi} |^2 + \hat{U}_\omega(\bar{\phi}) \right) + \omega Q^\prime \nonumber \\
&= S_3[\bar{\phi}(x)] + \omega Q^\prime, \label{eq:EnergyTunnelling}
\end{align}
where the first term is the three-dimensional Euclidean action of the ``bounce" solution tunneling between the two minima of the effective potential $\hat{U}_\omega(\phi) = U(\phi) - \omega^2 \phi^2 \slash 2$.

\subsection{Thin Wall Regime}

In the thin wall regime, the energy may be calculated in a different manner, due to~\cite{Spector}.  Beginning again with equation \eqref{eq:EnergyBasic}, we use the oscillatory time dependence to write:
\begin{align}
E(Q^\prime) &= \dfrac{Q^{\prime \, 2}}{2 \int \bar{\phi}^2 \, d^3x} + \int \dfrac{1}{2} \left( \nabla \bar{\phi} \right)^2 \, d^3x + \int U(\bar{\phi}) \, d^3x \\
&= \dfrac{Q^{\prime \, 2}}{2 \int \bar{\phi}^2 \, d^3x} + T + V,
\end{align}
where $T = \int (\nabla \bar{\phi})^2 \; d^3x\slash 2 $ and $V = \int U(\bar{\phi}) \, d^3x$.

In thin wall approximation, $ \phi \approx \phi_0 $ for $ r<R - \delta \slash 2 $ and $\phi \approx 0$ for $ r > R + \delta \slash 2 $, where $ \delta $ is the width of the surface of the Q-ball.  The thin wall approximation is valid if $ \delta \ll R $.  The ``volume" term then has two contributions, one from the interior of the Q-ball and one from the surface.  In the interior of the Q-ball, the potential is $U(\phi_0)$, while in the surface of the Q-ball, it is $ \beta m_0^2 \phi_0^2 $, where $\beta$ is a positive constant.  Therefore the ``volume" term is:
\begin{equation}
 V = \int U(\bar{\phi}) \, d^3x  = \dfrac{4}{3} \pi U(\phi_0) R^3 + 4 \pi \delta R^2 \cdot \beta m_0^2 \phi_0^2.
 \end{equation}

In the surface, the field changes by $\Delta \phi = \phi_0$ in the distance $\Delta r = \delta$, thus $d\phi \slash dr \approx \phi_0 \slash \delta$.  Introducing a constant $\alpha$ to account for the uncertainty, the surface term is:
\begin{equation}
T = 4 \pi \delta R^2 \cdot \alpha \dfrac{\phi_0^2}{\delta^2}.
\end{equation}

The energy must be a minimum with respect to both $ \delta $ and $ R $; minimizing with respect to $\delta$ gives $ \delta = \sqrt{ \alpha \slash \beta m_0^2} $ and
\begin{equation}
E= \dfrac{3 Q^{\prime \, 2}}{8 \pi \phi_0^2 R^3} + 8 \pi m_0 \sqrt{ \alpha \beta} \cdot R^2 \phi_0^2 + \dfrac{4}{3} \pi U(\phi_0) R^3.
\end{equation}
By manipulating equation \eqref{eq:EnergyTunnelling}, one can relate $\sqrt{\alpha \beta}$ to the one-dimensional Euclidean action for the true potential:
\begin{equation}S_1 = \int_0^{\phi_0} \sqrt{ 2 U(\phi)} \, d\phi = 2 m_0 \sqrt{\alpha \beta} \, \phi_0^2, \end{equation}
which is related to the three-dimesional action through $S_3 = 4 \pi R^2 U(\phi_0) \slash 3 + 4 \pi R^2 S_1$~\cite{Linde}.

Minimizing the energy with respect to $R$ results in a constraint between the charge and the radius:
\begin{equation}
0 = - \dfrac{9 Q^{\prime \, 2}}{8 \pi \phi_0^2} + 16 \pi m_0 \sqrt{\alpha \beta} \phi_0^2 R^5 + 4 \pi U(\phi_0) R^6. \label{eq:Constraint}
\end{equation} 
If $U(\phi_0)>0$, the last term would dominate over the second term, and if we neglect the second term, we could solve for $R$ in terms of the charge.  This would give the familiar $R \propto Q^{\prime 1\slash 3}$ behavior~\cite{Coleman}.  However, we have $U(\phi_0)<0$, and so one cannot neglect the second term.  This sixth order equation has no closed form solution.

\subsection{Thick Wall Regime}

For the thick wall approximation, we consider equation \eqref{eq:EnergyTunnelling}.  As the charge becomes small, the frequency $\omega$ becomes large; then the true asymmetric minimum in $U_\omega(\phi)$ is significantly lower than the symmetric minimum; this is true independent of the sign of $U(\phi_0)$; in fact, since we have $U(\phi_0) < 0$, the second minimum (the true vacuum) is lower than the symmetric minimum even at $\omega = 0$.  Therefore, the Euclidean action $S_3$ is the same in both cases, and the relations between the energy, radius, and charge are unchanged.  These are~\cite{Kusenko Small Qballs}:
\begin{align}
E &= Q^\prime m_0 \left( 1 - \dfrac{\epsilon^2}{6} - \dfrac{\epsilon^4}{8} - \dots \right) \\
R^{-1} &= \epsilon m_0 \left( 1 + \dfrac{1}{2} \epsilon^2 + \dfrac{7}{8} \epsilon^4 + \dots \right)
\end{align}
where $ \epsilon = Q^\prime A^2 \slash 3 S_\psi m_0^2 $ and $S_\psi \approx 4.85$ was determined numerically.  This is valid when:
\begin{equation} Q^\prime \ll \dfrac{3 S_\psi m_0}{\sqrt{ \lambda} A} \qquad  Q^\prime < \dfrac{3 S_\psi m_0^2}{2 A^2}. \label{eq:ThickConstraints} \end{equation}

The thick wall approximation, like the thin wall approximation, neglects quantum corrections; thus it breaks down when these are large, which occurs around $Q^\prime \lesssim 7$~\cite{Graham}.

\subsection{Intermediate Regime}

There is a regime between where the thin wall approximation is valid and where the thick wall approximation is valid; unfortunately no general solutions are known in this regime.  Therefore, we use a linear interpolation between the two regimes.  We will need only the radius in terms of the charge $Q^\prime$, for which we use:
\begin{equation}R = \dfrac{Q^\prime-7}{Q^\prime} R_{thin} + \dfrac{7}{Q^\prime} R_{thick}.\end{equation}

We have already written $R_{thick}$ in terms of $Q^\prime$; however, as found above there is no closed-form equation for $R_{thin}$ in terms of $Q^\prime$.  Therefore, we use a numerical approximation for the thin wall radius at small charge of the form $R \approx a + b Q^{\prime \; 2\slash 5}$.  While this can only be justified by its numerical accuracy, it can be motivated by neglecting the third term in the constraint equation~\eqref{eq:Constraint}.  Then in the intermediate regime the linear interpolation becomes:
\begin{equation}
R = \dfrac{Q^\prime-7}{Q^\prime} \left( a + b Q^{\prime \; 2\slash 5} \right) + \dfrac{21 S_\psi}{Q^{\prime \; 2} A^2 } .\label{eq:IntermediateRadius}
\end{equation}

\subsection{Bethe-Salpeter Regime}

The stability of states with very low charge is vital to building critically sized Q-balls, but for these states, one cannot use the approximations already discussed because quantum effects are important.  At small charge, it is furthermore incorrect to think of the Q-balls as made of $\phi$ quanta; instead we should remember that they are bound states made of squarks and Higgs bosons.  We will do this until the bound states are large enough that quantum effects are negligible, which occurs at $Q^\prime = 7$~\cite{Graham}.

For these small Q-balls, we consider them as bound states of squarks exchanging light Higgs bosons; this is approximately described by the Bethe-Salpeter equation in the ladder approximation.  This neglects diagrams where the rungs of the ladder are crossed; therefore, we use an effective coupling $\tilde{A}$ which we tune to ensure that the energy at large charge matches the result from the thick wall approximation.  The Bethe-Salpeter equation is discussed in more detail in the Appendix.

In this section, we work with the number $\mathsf n = 3Q$ of squarks present in the state; at the end, we will relate this to the charge $Q^\prime$ that we have been using by $Q^\prime = 3 Q \slash \sin^2(\theta) $, as in \eqref{eq:Charges}.  The lowest state is a single squark, and then the first step is the relatively simple case of two equal squarks forming one $\mathsf n=2$-ball; in this equal mass and equal coupling case the Bethe-Salpeter equation (after a Wick rotation) is:
\begin{align}
\left[ \left( \dfrac{M}{2} + p \right)^2 + m_q^2 \right] &\left[ \left( \dfrac{M}{2} - p \right)^2 + m_q^2 \right] \psi(p) \nonumber \\
&= \dfrac{\tilde{A}^{2}}{16 \pi^4} \int d^4k \dfrac{\psi(k)}{(p-k)^2 + m_h^2},
\end{align}
where $ \psi(p) $ is the wavefunction.  Approximating $ m_h=0 $, the bound state energies are~\cite{Bethe Salpeter-same}:
\begin{equation} M_n = 2 m \left( 1 - \dfrac{\alpha^2}{8 n^2} \right) = 2m_q \left( 1 - \dfrac{\tilde{A}^{4}}{2048 \pi^2 m_q^4 n^2} \right), \end{equation}
if $ \alpha = \tilde{A}^{2} \slash 16 \pi m_q^2 < 1 $.  We may find the binding energy for the ground state by taking $n=1$.

For the remaining states, the masses and couplings at the top and bottom of the ladder are unequal; after a Wick rotation, the Bethe-Salpeter equation in the ladder approximation is:
\begin{align}
\left[ (m+\Delta)^2 + (p-\imath \eta_1 P)^2 \right] &\left[ (m-\Delta)^2 +(p + \imath \eta_2 P)^2 \right] \psi(p) \nonumber \\
&= \dfrac{A^{\prime}}{\pi^2} \int d^4k \dfrac{\psi(k)}{(p-k)^2},
\end{align}
where the masses of the particles, $m_{top}$ and $m_{bottom}$, are $ m \pm \Delta $.  The coupling constant is $ A^\prime = g_{top} g_{bottom} \slash 16 \pi^2 = (\mathsf n-1)\tilde{A}^2 \slash 16 \pi^2 $, where $g_{top} = (\mathsf n-1) \tilde{A}$ is the coupling at the top of the ladder and $g_{bottom} = \tilde{A}$ is the coupling at the bottom of the ladder.  The total charge $Q$ of the resulting Q-ball is $ \mathsf{n} \slash 3$.  The energy-momentum four-vector of the bound state, $ P $, is given by $ (0,M) $ where $ M $ is the bound state mass.  $ \eta_1 $ and $ \eta_2 $ come from transforming to a ``center of momentum" reference frame; these are:
\[ \eta_1 = \dfrac{m_{top}}{m_{top} + m_{bottom}} \qquad \eta_2 = \dfrac{m_{bottom}}{m_{top} + m_{bottom}}. \]

The binding energies are~\cite{Bethe Salpeter}:
\begin{align}
M^2 &= 4 \Delta^2 + 4 m^2 \left( 1 - \dfrac{\Delta^2}{m^2} \right) \left( 1 - \dfrac{A^{\prime \, 2} \pi^2}{4 m^4} \dfrac{1}{(1-\Delta^2 \slash m^2)^2} \right),
\label{UnequalMasses}
\end{align}
as is explained in more detail in the Appendix.  We use this equation iteratively to find the masses and binding energies of the small charge Q-balls; the results are shown in Table~\ref{tb:BS}.
\begin{table}
\begin{ruledtabular}
\begin{tabular}{|c|c|c|}
\hline $ \mathsf n $ (No. Squarks) &Mass of Q-ball&Binding Energy \\ 
\hline 1 &$ m_q $&0 \\ 
\hline 2 &$ 2 m_q - .0000989 \tilde{A}^4 \slash m_q^3 $&$.0000989 \tilde{A}^4 \slash m_q^3$ \\ 
\hline 3 &$ 3 m_q - .0002309 \tilde{A}^4 \slash m_q^3 $&$.0001320 \tilde{A}^4 \slash m_q^3$\\ 
\hline 4 &$ 4 m_q - .0003793 \tilde{A}^4 \slash m_q^3 $&$.0001484 \tilde{A}^4 \slash m_q^3$\\ 
\hline 5 &$ 5 m_q - .0005376 \tilde{A}^4 \slash m_q^3 $&$.0001583 \tilde{A}^4 \slash m_q^3$\\ 
\hline 6 &$ 6 m_q - .0007025 \tilde{A}^4 \slash m_q^3 $&$.0001659 \tilde{A}^4 \slash m_q^3$\\ 
\hline 7 &$ 7 m_q - .0008721 \tilde{A}^4 \slash m_q^3 $&$.0001696 \tilde{A}^4 \slash m_q^3$\\ 
\hline 
\end{tabular} 
\end{ruledtabular}
\caption{Energies of small Q-balls from the Bethe-Salpeter equation.  We note that the dependence of the mass on $\tilde{A}$ matches the dependence of the mass in the thick wall regime on $A$, and similarly the dependence on $m_q$ matches the dependence in the thick wall regime on $m_0$.} \label{tb:BS}
\end{table}

We have calculated this until $\mathsf n = 7$, or $Q = 7 \slash 3$.  Using $\theta = 1.308$, this corresponds to $Q^\prime = 7.51$.  (Recall that we are measuring charge in units of the charge of $\phi$ and each squark has slightly more than 3 unit charges.)  Since this is greater than 7, the thick wall approximation is applicable~\cite{Graham}.  In the thick wall regime, a Q-ball with this charge has energy
\[ M = 7.51 m_0 - .333 A^4 \slash m_0^3. \]
The difference in the first terms comes from the fact that states described by the Bethe-Salpeter equation do not have exactly the same proportion of squarks and Higgs bosons as the Q-balls described in the thick wall regimes; however, the difference in these terms is 7.2 percent.  This can be further improved by moving the global minimum closer to the $\tilde{q}$ axis.  This is because:
\begin{equation}Q^\prime m_0 = \mathsf{n} \sqrt{m_{\tilde{q}}^2 \sin^2(\theta) + m_h^2 \cos^2(\theta) } \slash \sin^2(\theta), \end{equation}
as can be seen by comparing the potentials \eqref{eq:MSSMPotential} and \eqref{eq:Potential}, and using the relation between $\mathsf{n}$ and $Q^\prime$.

Comparing the second terms in the mass equations gives $\tilde{A} = 4.51 A$.  Using the value of $A$ in the potential in \eqref{eq:Potential} gives $\tilde{A} = 263$~GeV, while the value that we put into our original potential in \eqref{eq:MSSMPotential} is 240 GeV.  This is a significant difference, which we attribute to a combination of the inherent inaccuracy of the ladder approximation in the Bethe-Salpeter equation; it does not improve if we iterate further.

The equation for the masses given above is valid as long as $A^\prime \ll m^2$.   For $\mathsf{n}=7$, this is approximately $7 \tilde{A}^2 \slash 16 \pi^2 \cdot 4 m_q = .019$, and so the approximation remains valid.

\section{Critical Values For Phase Transition}

In the thin wall regime,  the interior of the Q-ball is in the true vacuum, which has negative energy density.  If charge continues to increase, the Q-ball expands, converting more of the space into the true vacuum.  At a particular value of the charge and radius, it expands uncontrollably, thereby converting all space into the true vacuum~\cite{Kusenko Solitosynthesis}.  This Q-ball induced phase transition can occur even when such a phase transition cannot be induced by thermal fluctuations.

As will be demonstrated in our numerical example, the critical charge is of order $10^5$, which is within the thin wall regime. At the critical point, not only is $dE \slash dR = 0$, but also $d^2 E \slash dR^2 = 0$, which gives the additional constraint
\begin{equation} 0 = \dfrac{9Q^{\prime \; 2}_c}{2 \pi \phi_0^2} + 16 \pi m_0 \sqrt{\alpha \beta} \phi_0^2 R_c^5 - 8 \pi U_0 R_c^6, \label{eq:Constraint2}
\end{equation}
where $ U(\phi_0)=-U_0 $ with $ U_0>0 $.  We solve the two constraint equations~\eqref{eq:Constraint} and~\eqref{eq:Constraint2} for the critical charge and critical radius:
\begin{align}
R_c &= \dfrac{10 m_0 \sqrt{\alpha \beta} \phi_0^2}{3 U_0} \\
Q^\prime_c &= 2 \pi \phi_0 \sqrt{\dfrac{8 U_0}{45} } \left( \dfrac{10 m_0 \sqrt{\alpha \beta}\phi_0^2 }{3 U_0} \right)^3.
\end{align}

\section{Solitosynthesis}

In thermal equilibrium, the number density of Q-balls of a particular charge is given by a Saha equation:
\begin{equation} n_{Q^\prime} = \dfrac{g_{Q^\prime}}{g_\phi g_{Q^\prime - 1}} n_\phi n_{Q^\prime - 1} \left( \dfrac{2\pi}{m_0 T} \right)^{3\slash 2}  e^(B_{Q^\prime} \slash T), \label{eq:Boltzmann}
\end{equation}
where $ B_{Q^\prime} $ is the binding energy of a soliton of charge $ Q^\prime $ and $ g_{Q^\prime} $ is the internal partition function of the soliton.  $ g_\phi $ is the number of degrees of freedom associated with the $ \phi $ field.  Noting that we chose to work with real fields, this is 3 from the color charge carried by $\phi$.  $n_\phi$, also called the charge density, is the number of free squarks; since $\theta \approx \pi \slash 2$, conceptually we can think of this as the number $Q^\prime = 1$-balls present.  This is given by:
\begin{align} n_\phi = \eta n_\gamma - \sum_{Q^\prime>2} Q^\prime n_{Q^\prime}, \end{align}
where the baryon asymmetry is $ \eta $ and in any radiation-dominated era the photon density is $    2 \zeta(3) T^3 \slash \pi^2 $~\cite{Weinberg}.

The typical approach would be to solve these coupled equations numerically.  However, the critical charge is generically of order $10^3$ to $10^5$, which leads to at least $10^3$ coupled equations.  It is infeasible to solve these simultaneously.  Therefore, we take a different approach following~\cite{Solitosynthesis} and consider the evolution of the single Q-ball.  One significant advantage is that we will see that the Q-ball grows fast enough that we can ignore charge depletion and set
\begin{equation}n_\phi \approx \eta n_\gamma = \eta \dfrac{2.404 T^3}{\pi^2}. \label{eq:Charge_Depletion} \end{equation}.

A single Q-ball grows or shrinks according to:
\begin{align}\dfrac{dQ^\prime}{dt} = r_{abs}(Q^\prime)-r_{evap}(Q^\prime),\end{align}
where $r_{abs}$ is the absorption rate and $r_{evap}$ is the evaporation rate.  By detailed balance, $n_{Q^\prime} r_{abs}(Q^\prime) = n_{Q^\prime+1} r_{evap}(Q^\prime+1)$; also, the rate of absorption is $r_{abs}= n_\phi v_\phi \sigma_{abs}(Q^\prime)$.  For large charges, $\sigma_{abs} \approx \pi R^2$.  We will see numerically that the radius does not change rapidly as a function of charge; then $\sigma_{abs}(Q^\prime) \approx \sigma_{abs}(Q^\prime-1)$.  These approximations give:
\begin{equation}
\dfrac{dQ^\prime}{dt} \approx n_\phi v_\phi \sigma_{abs}(Q^\prime) \left(1  - \dfrac{n_{Q^\prime-1}}{n_{Q^\prime}} \right). \label{eq:GrowthRate}
\end{equation}
Thus the determining factor is $n_{Q^\prime-1}\slash n_{Q^\prime}$: if it is less than one, absorption dominates, but if it is greater than one, evaporation dominates.  From the Saha equations~\eqref{eq:Boltzmann}, this important ratio is:
\begin{align}
\dfrac{n_{Q^\prime-1}}{n_{Q^\prime}} &= g_\phi \dfrac{\pi^2}{2.404 \eta}  \left( \dfrac{2\pi T}{m_0} \right)^{-3\slash 2} e^{-B_{Q^\prime} \slash T} \label{eq:Ratio}.
\end{align}
At large temperatures, the exponential is negligible and this scales as $T^{-3\slash 2}$.  However, this ratio is less than one only if $T > \eta^{-2\slash 3} m_0$, which is typically quite large, above the temperatures at which the supersymmetry which inspired our potential is typically broken.  Therefore, we expect evaporation to dominate at the temperatures when the Q-balls are formed.

As the temperature decreases, the exponential term is no longer negligible.  Because $-B_{Q^\prime} < 0$, this term decreases $n_{Q^\prime-1}\slash n_{Q^\prime}$.  Therefore, at some temperature $T_s$ absorption will dominate.  The ratio $n_{Q^\prime-1}\slash n_{Q^\prime}$ is equal to one at:
\begin{align} T_s = \dfrac{B_{Q^\prime}}{\ln(g_\phi) - \ln(\eta) + (3\slash 2) \ln(m_0 \slash T_s) - 1.34}. \end{align}

It is possible for the binding energy to be sufficiently large that solving this equation for temperature results in an imaginary value; returning to equation \eqref{eq:Ratio}, this happens when $B_{\tilde{Q}}$ is so large that the right hand side is always less than 1, which means that $n_{Q^\prime} > n_{Q^\prime - 1} $ always.  Physically, whenever a $(Q^\prime - 1)$-ball forms, it will always grow into a larger ball; we may say that the solitosynthesis temperature for these charges is infinite.

Smaller charges have smaller binding energies, and so we will find real solitosynthesis temperatures for small Q-balls.  Therefore, we will need to wait for these smaller Q-balls to form, and then wait for these to grow into the larger ones which can always grow.

As this suggests, $T_s$ is greater for larger charges, which we will verify numerically.  Therefore Q-ball growth is a winner-take-all-situation, and the solitosynthesis temperature cannot cut off a growing Q-ball.

\subsection{Rate of Diffusion}

A Q-ball grows by absorbing the nearby charge.  If the charge is not be replenished sufficiently quickly through diffusion, there may be a local depletion of charge near the Q-ball which limits its growth.  If this occurs, the rate of growth will be given by $r_{diff}$, the rate that free squarks diffuse into the surface of the Q-ball, instead of $r_{abs}$.

Reference~\cite{Diffusion} is concerned with the related process of the diffusion of evaporating squarks away from a Q-ball.  The particle flux through the Q-ball surface is given by:
\begin{equation} r_{diff} = \dfrac{dQ^\prime}{dt} = - 4\pi k R D n_\phi^{eq}, \end{equation}
where $ D \approx aT^{-1} $, $ a \approx 4 $ for relativistic squarks, and $ k \approx 1 $ was determined numerically.  We need to adjust this equation because we are concerned with particles diffusing towards the Q-ball; the rate has the opposite sign and we multiply this by the velocity of the squarks because they are moving non-relativistically.  Thus:
\begin{equation} r_{diff} = v_\phi 16 \pi R T^{-1} n_\phi^{eq}. \end{equation}

We wish to compare this to $r_{abs}$, the rate of absorption as approximated above; the ratio is: $ r_{diff} \slash r_{abs} = 4 T^{-1} \slash R $.  Perhaps surprisingly, this is small for high temperatures and large for low temperatures.  This occurs because the rate of diffusion is propotional to $ T^{5\slash 2}$ while the rate of absorption is proportional to $T^{7 \slash 2}$.  Even though diffusion is decreasing as the temperature decreases, the rate of absorption drops faster; therefore, diffusion will limit the growth of Q-balls for temperatures above $4 \slash R$.  For radii of order .01 inverse GeV to .1 inverse GeV, this temperature is of order 40 GeV to 400 GeV, which is significantly above the solitosynthesis temperatures.  Therefore, diffusion will replenish the charge sufficiently quickly at the relevant temperatures.

This, combined with the winner-take-all behavior, demonstrates that global depletion of charge is not an issue, provided that most of the charge is in free squarks during solitosynthesis.  We will verify this numerically.

\subsection{Rate of Growth in the Thin Wall Regime}

Next we consider the rate of growth of the Q-balls in the various regimes.  For temperatures below the solitsynthesis temperature, the rate of evaporation is small, and we may approximate $ dQ^\prime \slash dt = n_\phi v_\phi \sigma_{abs}(Q^\prime) $ from equation~\eqref{eq:GrowthRate}.  Since charge depletion is negligible, we may assume $ n_\phi = \eta n_\gamma $.  The $Q^\prime=1$-balls being absorbed are in thermal equilibrium at $T \ll m_0$ with average velocity $ v_\phi = \sqrt{ 2T \slash \pi m_0 } $.  Additionally, we use the geometric area $\pi R^2 $ for the cross section.
In the radiation-dominated era, the temperature and the time are not independent; they are related by~\cite{Weinberg}:
\begin{equation}
t = \dfrac{1}{T^2} \sqrt{ \dfrac{3}{16 \pi G \mathcal{N}}} + \mathrm{constant}, \end{equation}
where $\mathcal{N}$ is the effective number of degrees of freedom of the particles in thermal equilibrium, with fermionic degrees of freedom weighted by $7 \slash 8$.  In our toy model with only the Higgs boson, squarks, and photons, we have $\mathcal{N}=6$.  Then we have:
\begin{equation}  dt = - 1.34 \cdot 10^{18} \; \mathrm{GeV} \, dT \slash T^3.\end{equation}
Thus our differential equation is:
\begin{equation} - \dfrac{1}{1.34 \cdot 10^{18} \; \mathrm{GeV}} \dfrac{dQ^\prime}{dT} = \pi R^2 \eta \dfrac{2.4}{\pi^2} \sqrt{ \dfrac{2T}{\pi m_0} } \label{eq:DEgrowth}, \end{equation}
using $ n_\gamma = 2.4 T^3 \slash \pi^2 $.

The right-hand side involves the radius which is not independent of the charge; however, in the thin wall approximation the radius cannot be written in terms of the charge because of the form of the 6th order equation relating them.  Fortunately, one can write the charge in terms of the radius, and then we consider the rate of the growth of the radius of the Q-ball until it reaches the critical radius:
\begin{align}
Q^\prime &= \sqrt{ \dfrac{128 \pi^2 m_0 \sqrt{\alpha \beta} \phi_0^4}{9} R^5 - \dfrac{32 \pi^2 U_0 \phi_0^2}{9} R^6 }  \nonumber \\
&\equiv \sqrt{ a_5 R^5 - a_6 R^6}.
\end{align}
Then the differential equation is:
\begin{equation} R^{-1\slash 2} \dfrac{5 a_5 + 6 a_6 R}{2\sqrt{ a_5 + a_6 R}} \; dR = - \dfrac{8.17 \cdot 10^{17} \; \mathrm{GeV} \eta}{\sqrt{m_0}}  T^{1\slash 2} \, dT. \end{equation}

Both sides of this equation can be integrated explicitly:
\begin{widetext}
\begin{align}
&3 \sqrt{a_5 R + a_6 R^2} + \dfrac{a_5}{\sqrt{ a_6 }} \ln \left( \dfrac{a_5 + 2 a_6 R + 2\sqrt{ a_6 R \left( a_5 + a_6 R \right) } }{2 \sqrt{a_6}} \right) - 3 \sqrt{a_5 R_i + a_6 R_i^2} \nonumber \\
&- \dfrac{a_5}{\sqrt{ a_6}} \ln \left( \dfrac{a_5 + 2 a_6 R_i + 2\sqrt{ a_6 R_i \left( a_5 + a_6 R_i \right) } }{2 \sqrt{a_6}} \right) = \dfrac{2}{3} \cdot \dfrac{8.17 \cdot 10^{17} \eta  \; \mathrm{GeV}}{\sqrt{m_0}} \left( T_{start}^{3\slash 2} - T^{3\slash 2} \right), \label{eq:ThinWallGrowth}
\end{align}
\end{widetext}
where $R_i$ is the radius of the smallest Q-ball at which the thin wall approximation is valid and $T_{start}$ is the temperature at which this Q-ball starts to grow.  This can be less than $T_s$ if these Q-balls do not form until a lower temperature.  If we set $R=R_c$, this equation can be solved for the temperature at which the Q-ball becomes critically sized.

\subsection{Rate of Growth in the Thick Wall Regime}

We begin with the differential equation \eqref{eq:DEgrowth} which is valid in the thick wall regime also.  We directly relate the radius to the charge, $ R = 3 S_\psi m_0 \slash Q^\prime A^2 $; then the differential equation becomes:
\begin{equation} \dfrac{dQ^\prime}{dT} = - 7.36 \cdot 10^{18} \; \mathrm{GeV} \dfrac{S_\psi^2 m_0^{3\slash 2} \eta}{Q^{\prime \, 2} A^4} \sqrt{T}, \end{equation}
whose solution is:
\begin{equation} Q^{\prime \; 3}_f - Q^{\prime \; 3}_i = 14.7 \cdot 10^{18} \; \mathrm{GeV} \dfrac{S_\psi^2 m_0^{3\slash 2} \eta}{A^4} \left( T_{start}^{3\slash 2} - T_f^{3\slash 2} \right), \end{equation}
where $T_{start}$ is the starting temperature for thick wall growth.  This is either the $Q^\prime=7$ solitosynthesis temperature, or the temperature at which $Q^\prime = 7$-balls form, whichever is smaller.

\subsection{Rate of Growth in the Intermediate Regime}

\paragraph{}We again begin with the differential equation \eqref{eq:DEgrowth} and use the linear interpolation for the radius, equation \eqref{eq:IntermediateRadius}, which gives:
\begin{align}
\int_7^{Q^\prime_f} &\dfrac{Q^{\prime \; 2} dQ^\prime}{\left( (Q^\prime -7) (a + b Q^{\prime \; 2 \slash 5} ) + 21 S_\psi \slash Q^{\prime \; 2} A^2 \right)^2} \nonumber \\
&= \dfrac{5.45 \cdot 10^{17} \; \mathrm{GeV} \eta}{\sqrt{m_0}} \left( T_{start}^{3\slash 2} - T_f^{3\slash 2} \right) .\label{eq:InterGrowth}
\end{align}
The left hand side of this equation must be integrated numerically.

\subsection{Bethe-Salpter Equation Regime}

We next consider the growth of very small Q-balls; in this regime, cross sections cannot be approximated by the geometrical area and so equation \eqref{eq:DEgrowth} is not valid.  However, because these are the first steps of solitosynthesis, all of the charge will be in these lowest seven states, which we label with $\mathsf n$, the number of squarks present in the state.  Therefore, one can return to the initial method of considering the evolution of the number densities as a function of temperature; we have 8 equations to solve numerically, instead of $10^5$.

The number densities of the Q-balls are given by the Saha equations like \eqref{eq:Boltzmann}, which we write in terms of fractional densities $X_{\mathsf n} = n_{\mathsf n}\mathsf n \slash N $, where $N$ is the total number of squarks, $\eta \cdot 2.404 T^3 \slash \pi^2$:
\begin{align}
X_{\mathsf{n}} &= \dfrac{\mathsf{n}}{\mathsf{n}-1} \dfrac{2.404 \eta}{3 \pi^2} \left( \dfrac{2\pi T}{m_0} \right)^{3\slash 2} X_{\mathsf{n}-1} X_1 e^{B_\mathsf{n}\slash T},
\label{eq:FracDen}
\end{align}
with the additional equation $X_1 + X_2 + X_3 + X_4 + X_5 + X_6 + X_7 = 1$.  (The 3 in the denominator of the Saha equations comes from the 3 color degrees of freedom for a real squark field.)

We observe that we do not need a generic Q-ball to grow into a critically-sized Q-ball to induce the phase transition, but only one per Hubble volume, $1 \slash H^3$ where $H= T^2 \slash 2.43 \cdot 10^{18} \; \mathrm{GeV}$.  Therefore, our generic approach is to find temperature at which there are of order 10 $\mathsf{n}=7$-balls per Hubble volume, which must be done numerically.  Since $Q^\prime =7.25$ when $\mathsf{n}=7$, we may begin the thick wall analysis once sufficiently many $\mathsf{n}=7$-balls form.  We note that we must also verify that at this temperature most of the charge remains in $\mathsf{n}=1$-balls; otherwise our analysis above is invalid because we neglected charge depletion.


We see that the exponential suggests that $X_7$ will become large at low temperature, and due to the $T^{3\slash 2}$, we also expect $X_7$ to be large at higher temperatures.  Thus, we generically expect $X_7$ to grow at both large and small temperatures, with a minimum between.  Typically, the solitosynthesis temperature for $Q^\prime = 7$-balls is after the number of $\mathsf{n}=7$-balls per Hubble volume has dropped beneath 1; that is, most of the $\mathsf{n}=7$-balls have evaporated away.  Then we need to wait until $\mathsf{n} = 7$-balls form again at lower temperatures.

However, there exist cases in which the number of $\mathsf{n} = 7$-balls per Hubble volume is still greater than 1 when the universe cools to the solitosynthesis temperature of $Q^\prime = 7$-balls; in this case, they may begin to grow immediately.  In fact, then our analysis underestimates the temperature at which the phase transition occurs.  Since larger Q-balls begin accreting charge earlier, it is likely that there are even larger Q-balls that have not evaporated away when the temperature reaches their (higher) solitosynthesis temperature.  Again, though, we reiterate that our goal is simply to investigate the theoretial existence of such a phase transition.  If one of these $\mathsf{n}=7$-balls has time to induce a phase transition, then we can be certain that any larger ones that had the opportunity to grow earlier would also induce a phase transition, and thus we still conclude that the phase transition does occur.

It should be noted that especially in such a case it is important to verify that most of the charge is in $\mathsf{n}$=1-balls before we can use the results derived above; this means solving equations \eqref{eq:FracDen} numerically.

\section{Freeze Out}

Q-ball growth can be ended in one of two ways: either the necessary reactions freeze out as the universe expands, or the Q-balls deplete the nearby charge.  We have already demonstrated that  charge depletion does not hinder the growth of at least one critically sized Q-balls per Hubble volume; therefore we need only to consider freeze out.

The reactions responsible for Q-ball growth freeze out when their time scale is greater than the Hubble time scale, $ \tau_H = H^{-1} $.  While the universe is radiation dominated, the Hubble constant $T^2 \slash M_{Pl}$, and so the Hubble time scale is $ \tau_H = \mathrm{2.43 \cdot 10^{18} \; GeV}\slash T^2 $.

The time scale of Q-ball growth is $\tau_{abs} = 1\slash r_{abs} =  1\slash n_\phi \sigma v_\phi $.  We consider the later reactions in the sequence; then the heavy Q-balls are effectively at rest and the $Q^\prime=1$-balls are moving non-relativistically in thermal equilibrium, with $v_\phi = (2 \slash \sqrt{\pi}) \sqrt{2T \slash m_0} $.  We use the geometric cross section, $ \sigma = \pi R^2 $; then 
\begin{equation}\tau_{abs} = \dfrac{\pi^2}{4.8 \eta T^3 R^2} \sqrt{ \dfrac{2T}{m_0} }. \end{equation}

Setting these timescales equal and solving for $T$ gives
\begin{equation} T = \left( \dfrac{\sqrt{2} \pi^2}{2.43 \cdot 10^{18} \; \mathrm{GeV} \cdot 4.8 \eta} \cdot \dfrac{1}{R^2\sqrt{m_0}} \right)^2.\end{equation}

When we do our numerical analysis below, we will find that these are orders of magnitude smaller than the temperatures relevant to the phase transition.

\section{Numerical Example} 

Finally, we demonstrate that the potential above is one in which all of these processes work out.  As an reminder, the numbers above give $m_0 =193 \; \mathrm{GeV}$, $A = 58.2 \; \mathrm{GeV}$, and $\lambda = .0450$, which gives a potential where the thin wall approximation is valid for large charge.  Above we also found that the minimizing field is $\phi_0 = 646 \; \mathrm{GeV}$ at which the potential is $-U_0$ with $U_0 = 6.18 \cdot 10^7 \; \mathrm{GeV}^4$.

The other constant that must be set is $\eta$, the baryon asymmetry.  In the actual universe, this is about $5 \cdot 10^{-10}$.  Again, though, we emphasize that our goal is to demonstrate that this phase transition is a theoretical possibility, and not necessarily part of the evolution of the universe.  Therefore, in this first numerical example, we will take $\eta = 3 \cdot 10^{-6}$.  This value illustrates the phase transition well, although the final temperature will dip below the QCD confinement temperature; therefore we will also give a second, although more complicated, numerical example which avoids this.

As regards the phase transition, the critical charge is $5.36 \cdot 10^5$ and the critical radius is $.341 \; \mathrm{GeV}^{-1}$.  We present a table of the radii and solitosynthesis temperatures for various charges in Table \ref{tb:TS}.  With these radii and our chosen value of $\eta$, the freeze-out temperature is of the order $10^{-12}$~GeV, which is significantly smaller than any of the temperatures we will consider.
\begin{table}
\begin{ruledtabular}
\begin{tabular}{|c|c|c|c|}
\hline 
Charge $Q^\prime$ & Radius ($\mathrm{GeV}^{-1}$) & $ B_{Q^\prime} $ ($\mathrm{GeV}$) & $T_s \; (\mathrm{GeV})$ \\ 
\hline 
$ Q_c = 5.36 \cdot 10^5 $ & .341 & $9.31 \cdot 10^7 $ & $\infty$ \\ 
\hline 
5000 & .0375 & $6.20 \cdot 10^5 $ & $\infty$ \\ 
\hline 
1000 & .0195 & $9.60 \cdot 10^4 $ &  16600 \\ 
\hline 
500 & .0148 & $4.05 \cdot 10^4 $ & 5420 \\ 
\hline 
200 & .0102 & $1.16 \cdot 10^4 $ & 1190 \\ 
\hline 
7 & .118 & .482 & .0163 \\ 
\hline 
\end{tabular} 
\end{ruledtabular}
\caption{Solitosynthesis temperatures for several charge values.  By infinity, we mean that such a Q-ball always grows.  The top three were calculated in the thin wall regime, while the last one was calculated in the thick wall approximation.  The other two are technically in the intermediate regime; to approximate their solitosynthesis temperatures we used the thin wall regime.  The important point is that because the temperature rise, they cannot cutoff a growing Q-ball.} \label{tb:TS}
\end{table}

\subsection{Bethe-Salpeter Growth}

We begin with solitosynthesis in the Bethe-Salpeter regime.  At the $Q^\prime = 7$ solitosynthesis temperature, there are of order $10^{-6}$ $\mathsf{n}=7$-balls per Hubble volume.  Therefore, we need to wait until these small Q-balls form again at low temperatures before thick wall growth can begin.  Numerically, we find that there are order 10 $\mathsf{n} =7$-balls at $T = .00889 \; \mathrm{GeV}$.

\begin{figure*}
\includegraphics[scale=.85]{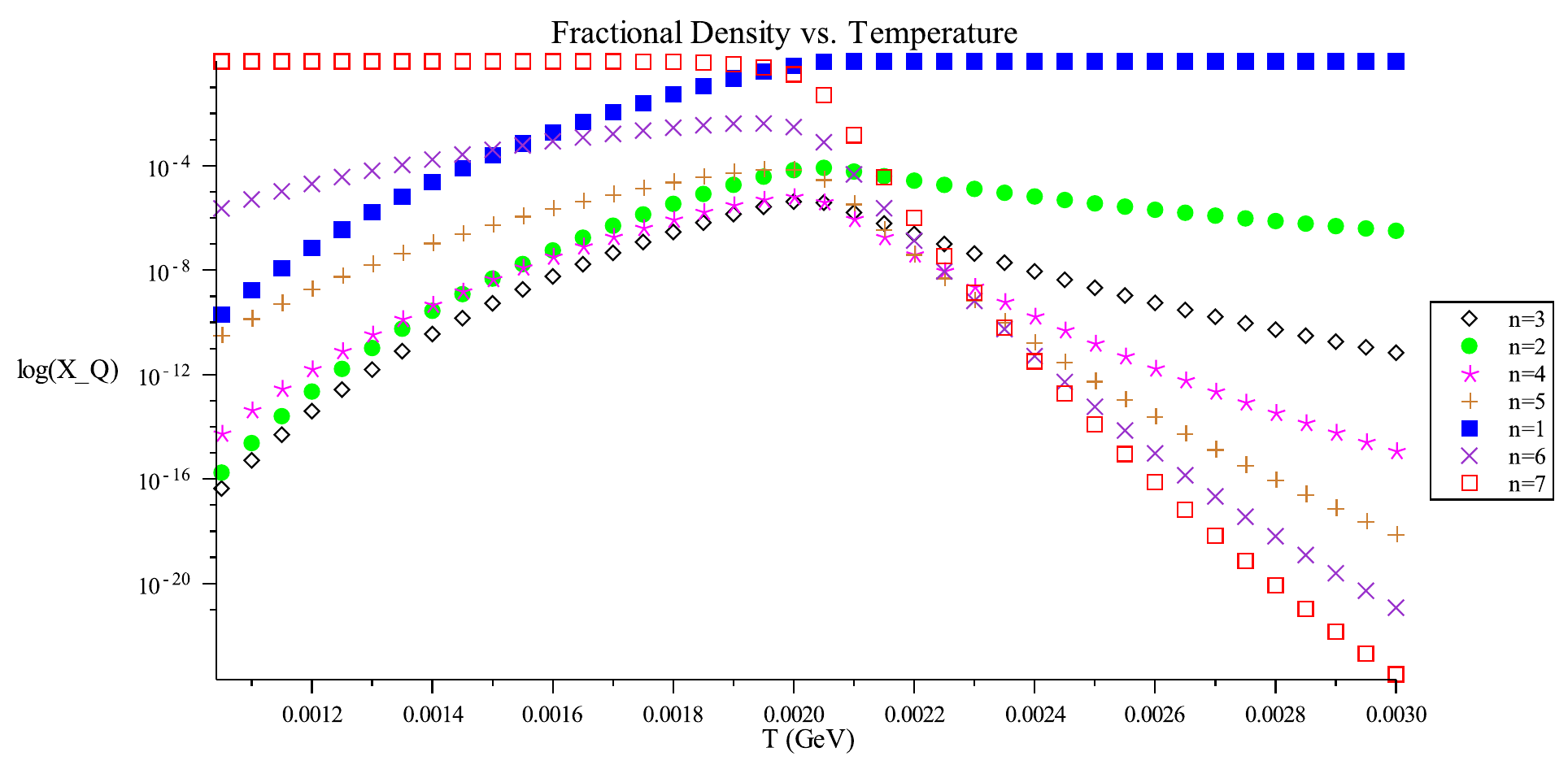}
\caption{The growth of very small Q-balls.  $X_{\mathsf{n}}$ is the charge density of each type of Q-ball, given by Eq~\eqref{eq:FracDen}.  Notice that as temperature decreases, the number of $\mathsf{n} = 7$-balls increases, while the number of $\mathsf{n}=1$-balls decreases, as we expect.} \label{fig:BSGrowth}
\end{figure*}

Next we must address charge depletion.  At the starting temperature of $.00889$~GeV, over 99.9999999 percent of the charge is in individual squarks, and so we are justified in ignoring charge depletion.  However, we also need to know what is the lowest temperature for which this assumption is valid; if our final temperature is beneath this, then our analysis is untrustworthy.  To determine this bound, we consider at what temperature the majority of the charge is no longer in $\mathsf{n}=1$-balls (which are individual squarks), if we ignore all of the states above $\mathsf{n}=7$.  These charge densities are shown in Figure~\ref{fig:BSGrowth}.  We see that the majority of the charge is no longer in individual squarks around $T = .0019$~GeV.  As long as our final temperature is above this, we are justified in neglecting charge depletion.

\subsection{Thick Wall and Intermediate Growth}

The thick wall approximation is valid until $Q^\prime = 80$; the two constraints \eqref{eq:ThickConstraints} give $Q^\prime < 80$ and $Q^\prime < 228$.  The growth in this regime is virtually instantaneous; the temperature drops by less than one part in $10^{9}$.  Thus, the starting temperature is still $.00889$~GeV for growth in the intermediate regime.

\begin{figure}
\includegraphics[scale=.45]{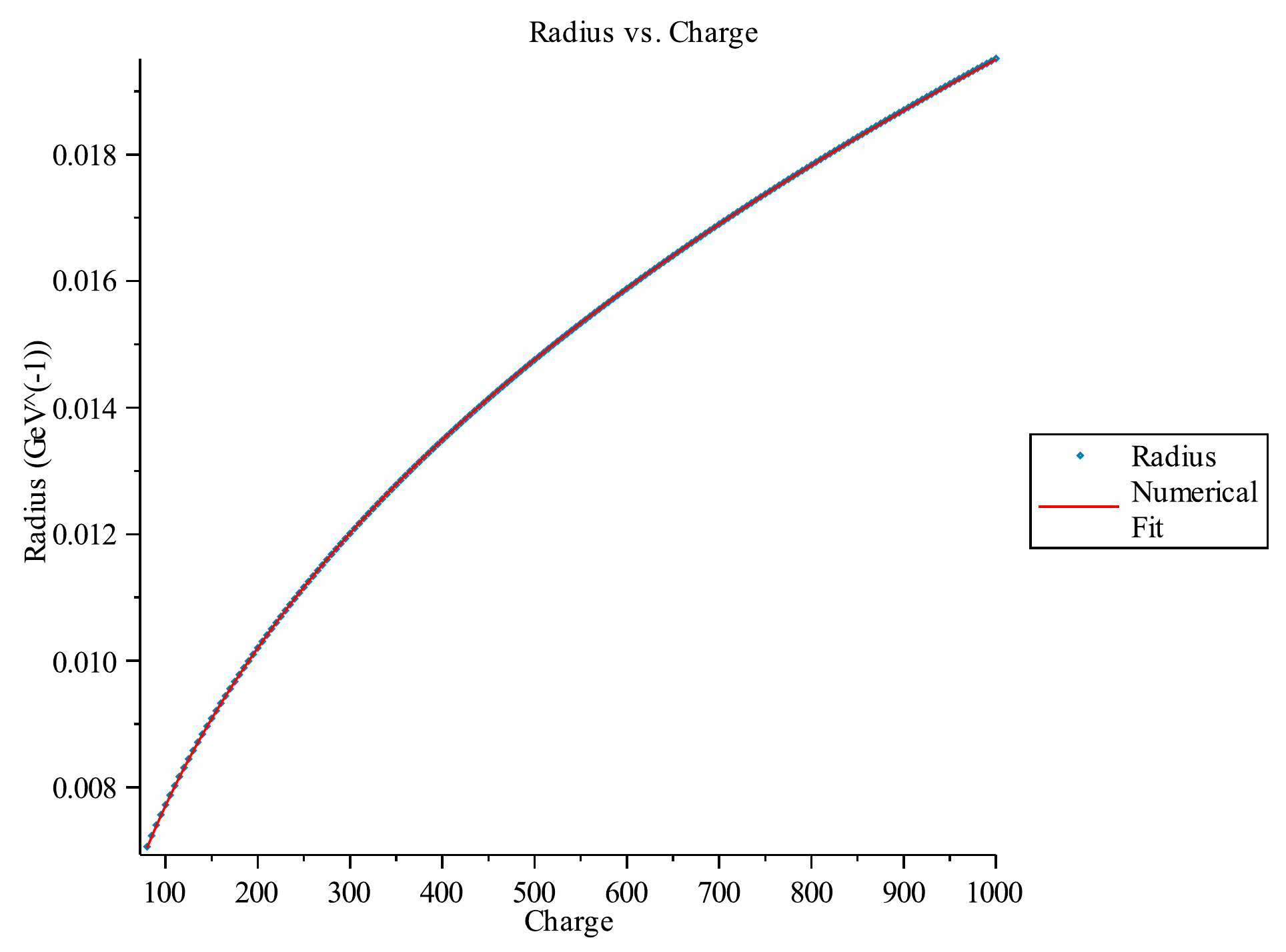}
\caption{The numerical fit for the radius as a function of charge $Q^\prime$ in the thin wall approximation, for small charges.  This is the fit to be used in the interpolation for the intermediate regime.} \label{fig:LinFit}
\end{figure}

The thin wall regime becomes applicable for $T \gg 228$; therefore we will use the intermediate regime for charges between 80 and 1000.  First, however, we must find the constants $a$ and $b$.  We numerically fit the function $a + b Q^{\prime \; 2\slash 3}$ to the radius for small values of $Q^\prime$; the result is plotted in Figure~\ref{fig:LinFit}.  This fit gives $a = -9.21 \cdot 10^{-5}$ and $ b = 1.24 \cdot 10^{-3} $.  Substituting this into differential equation \eqref{eq:InterGrowth} and solving numerically gives us $T_f = .00855$~GeV for the temperature what the Q-ball reaches $Q^\prime = 1000$.  This is, of course, less than the solitosynthesis temperature for such a Q-ball, so thin wall growth begins immediately.

\subsection{Thin Wall Growth}

\begin{figure}
\includegraphics[scale=.42]{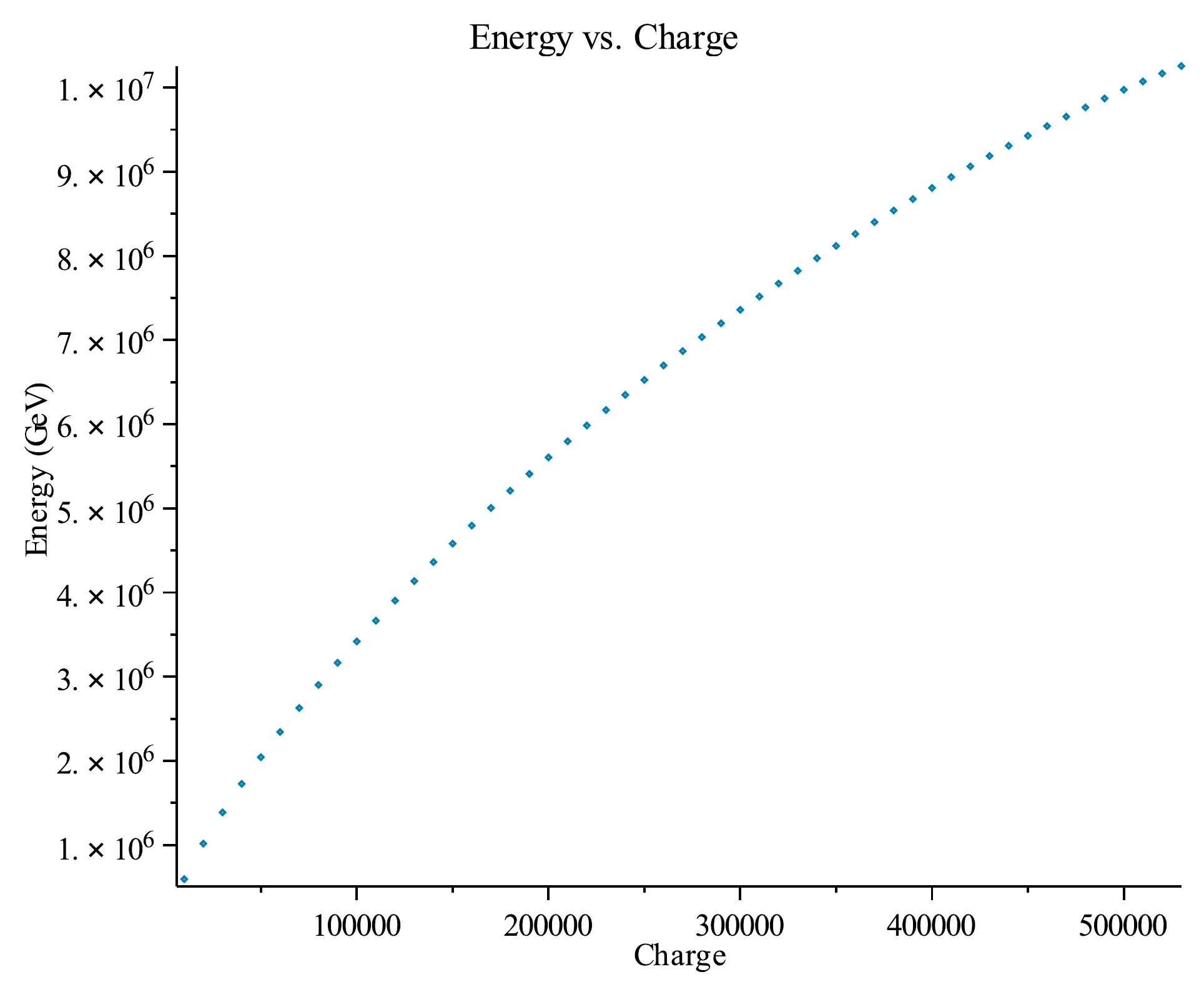} 
\includegraphics[scale=.42]{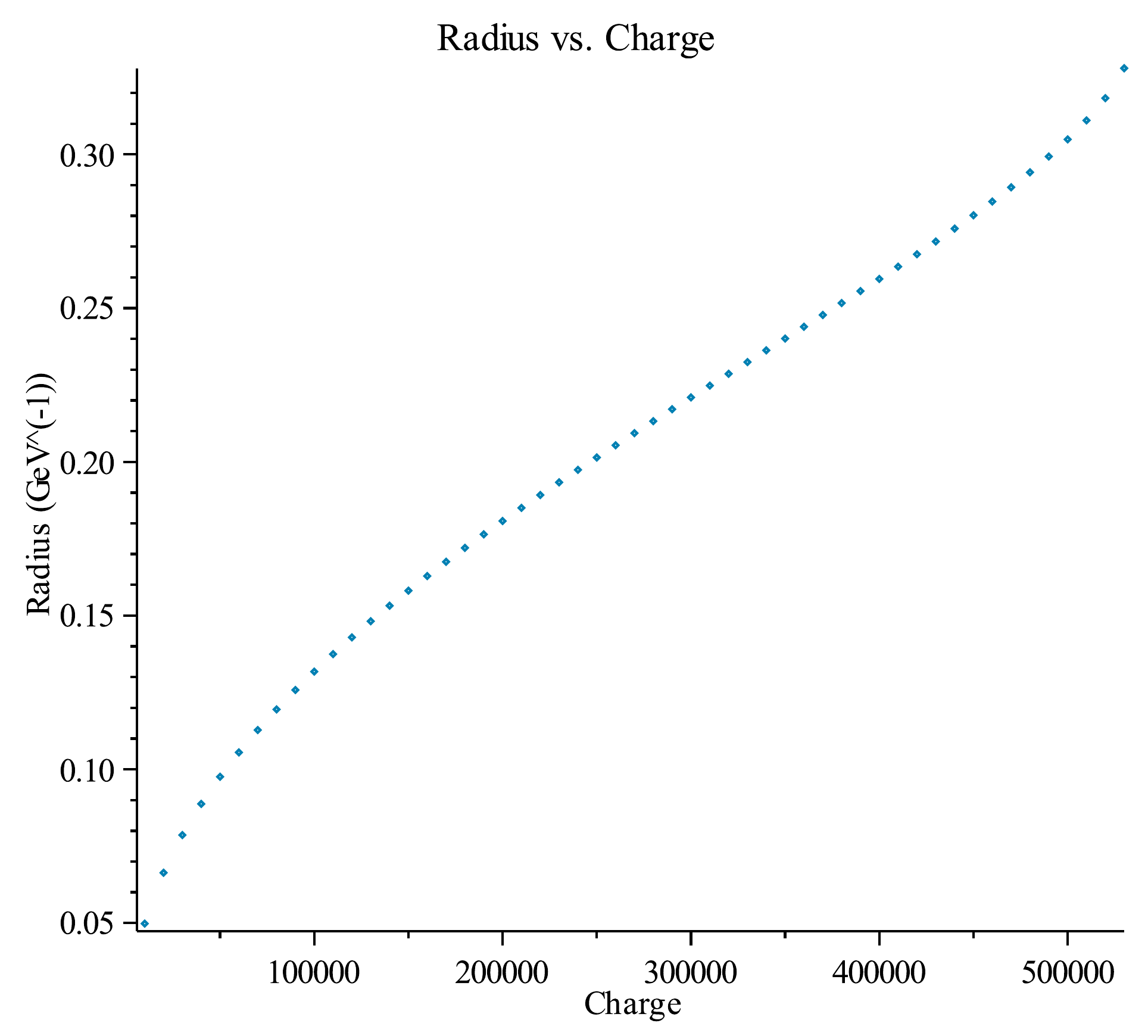}
\caption{The energy and radius as a function of charge in the thin wall regime, calculated numerically.} \label{fig:ThinPlots}
\end{figure}

First, we show the energy and radius for the thin wall approximation; these are shown in Figure \ref{fig:ThinPlots}.  When we derived the solitosynthesis temperature, we assumed that $R_{Q+1} \approx R_{Q}$; we see that throughout the thin wall regime this is justified.
Using $T= .00855 \; \mathrm{GeV}$ as the starting temperature in equation \eqref{eq:ThinWallGrowth}, we find that the Q-ball grows to critical size at $T = .00493 \; \mathrm{GeV}$.  This temperature is greater than $.0019$~GeV, and furthermore, it is greater than the freeze-out temperature scale; therefore we conclude that such phase transitions are indeed possible.

\section{A Second Numerical Example}

As mentioned, the previous numerical example is not strictly speaking acceptable, because the final temperature is less than the QCD scale, $\Lambda = .217$~GeV, at which confinement introduces additional complications.  Therefore, we consider a second example, which avoids this problem.  We use the same potential as above; thus $m_0$, $A$, and $\lambda$ are unchanged.  The critical charge and radius are also unchanged, as are the ranges where the thick and thin wall approximations are applicable.  Similarly, the numerical fit for the radius at small charges used in the intermediate regime is unchanged.

\begin{figure}
\includegraphics[scale=.4]{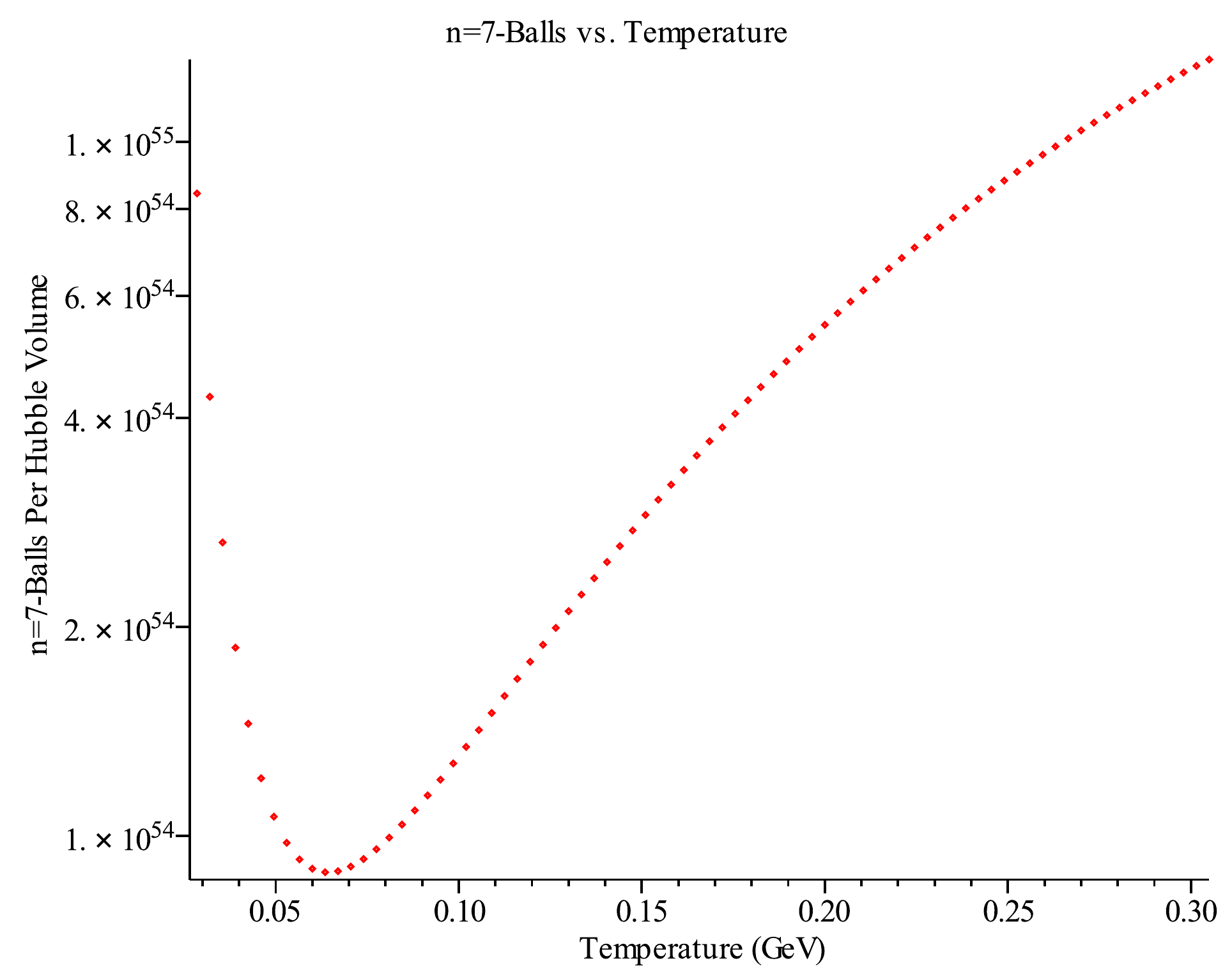}
\caption{The number of $\mathsf{n}=7$-balls per Hubble volume.  As expected, the number rises at both high and low temperatures.  At $T = .258$~GeV, there are sufficiently many for thick wall growth to begin, even though they are still evaporating away into small Q-balls.} \label{fig:n7perHubble}
\end{figure}

We will, however, choose an exceptionally large $\eta$, of the size of $3 \cdot 10^3$.  Then the solitosynthesis temperature for $Q^\prime=7$-balls is $.258$~GeV, above the QCD scale.  Furthermore, at this temperature, there are still of order $10^{54}$ $\mathsf{n}=7$-balls per Hubble volume; thick wall growth may begin immediately.  This is one of the exceptional cases mentioned in the discussion on growth in the Bethe-Salpeter regime; even though these Q-balls are evaporating away, there are sufficiently many of them for growth to begin immediately.  For clarity, we plot the number density of $\mathsf{n}=7$-balls per Hubble volume in Fig. \ref{fig:n7perHubble} to show it has the expected behavior.  Furthermore, at this temperature, 71 percent of the charge is in individual squarks, and so we may ignore the charge depletion if the phase transition occurs sufficiently rapidly.

We will show that this Q-ball can grow into a critically charged Q-ball and induce the desired phase transition.  However, as discussed in the section on growth the Bethe-Salpeter regime, this may underestimate the temperature as which the phase transition occurs, because larger Q-balls may have evaporated away sufficiently slowly that there are still sufficiently many of them to induce a phase transition when they begin accreting charge at their larger solitosynthesis temperature.  Since there are so many $\mathsf{n}=7$-balls at their solitosynthesis temperature, this is extremely likely.  However, if these $\mathsf{n}=7$-balls can induce a phase transition, then we can be certain that any larger ones that began growing earlier would also induce the phase transition, and so the phase transition certainly occurs. 

As above, the thick wall growth is virtually instantaneous; there is no appreciable drop in the temperature.  Similarly, the growth in the intermediate regime is also extremely fast; the temperature drops less than 1 part in $10^9$.  Finally, growth in the thin wall approximation is equally fast; again the temperature changes by less than one part in $10^9$.  Thus, such a Q-ball becomes critically charged within a temperature change of .000000001 GeV, and so the final temperature is still above the QCD scale.

However, we can merely conclude that the phase transition does occur; we cannot conclude that this is the temperature it occurs at, as we could above.  As mentioned, it is very likely that it in fact occurs at a larger temperature.  This is why we gave the first numerical example, which is a more typical case.

\section{Potential MSSM Applications}

While the theoretical possibility of such a phase transition is in itself intersting, one would also like to know whether such a phase transition could occur in extensions of the Standard Model such as supersymmetry, which naturally provides squarks carrying baryon number.  This analysis suggests that, provided that the requisite vacuum structure can be found, such phase transitions are indeed possible.

Indeed, one can ask whether such a phase transition is possible within the evolution of our own universe.  If squarks do exist, they must be significantly heavier than quarks.  Therefore they will decay rapidly, and thus we cannot build critically charged Q-balls out of them in our current vacuum.  However, such a phase transition could have occurred in the past, if the vacuum structure has these requisite properties:
\begin{enumerate}
\item A global minimum in which no squarks or sleptons develop vacuum expectation values, so that baryon number and lepton number are conserved.
\item A local minimum in which no squarks develop vacuum expectation values, so that baryon number is conserved.
\item In the local minimum, quarks must be heavier than squarks, so that they are stable against decay into quarks.
\item In the local minimum, one of the bosons that mediates an interaction between squarks must be lighter than the squarks; this is required for bound states to develop.
\item The potential expanded in the local minimum must allow the creation of Q-balls through the squark fields.
\item The barrier between the local minimum and the global minimum must be sufficiently large to suppress tunneling between the minima by thermal fluctuations.
\end{enumerate}

Such a vacuum can indeed by found; as an example, consider an MSSM potential of the form:
\begin{widetext}
\begin{align}
U &= -m_h^2 H^* H + m_{\tilde{Q}}^2 \tilde{Q}^* \tilde{Q} + m_{\tilde{q}}^2 \tilde{q}^* \tilde{q} + m_{\tilde{L}}^2 \tilde{L}^* \tilde{L} + m_{\tilde{l}}^2 \tilde{\ell}^* \tilde{\ell} + \dfrac{\lambda}{4} (H^* H)^2  -A_S \left( H \tilde{Q}^* \tilde{q} + H^* \tilde{Q} \tilde{q}^* \right)- A_L \left( H \tilde{L}^* \tilde{l} + H^* \tilde{L} \tilde{l}^* \right) \nonumber \\
&\qquad + y^2 \left( H^* H \tilde{Q}^* \tilde{Q} + H^* H \tilde{q}^* \tilde{q} + \tilde{Q}^* \tilde{Q} \tilde{q}^* \tilde{q} \right) + y^2 \left( H^* H \tilde{L}^* \tilde{L} + H^* H \tilde{l}^* \tilde{l} + \tilde{L}^* \tilde{L} \tilde{l}^* \tilde{l} \right)+ \dfrac{g_1^2}{8} \left( H^* H - \tilde{Q}^* \tilde{Q} \right)^2  \nonumber \\
&\qquad + \dfrac{g_1^2}{8} \left( H^* H - \tilde{L}^* \tilde{L} \right)^2  + \dfrac{g_2^2}{8} \left( H^* H + \tilde{Q}^* \tilde{Q} - 2 \tilde{q}^* \tilde{q} \right)^2  + \dfrac{g_2^2}{8} \left( H^* H + \tilde{L}^* \tilde{L} - 2 \tilde{l}^* \tilde{l} \right)^2,
\label{eq:MSSM_Potential}
\end{align}
\end{widetext}
where $\tilde{Q}$ and $\tilde{q}$ are squarks, $\tilde{L}$ and $\tilde{\ell}$ are sleptons, and $H$ is a Higgs boson, although again for simplicity, we take these fields to be real.  One local minimum of this potential is at $\left< \tilde{Q} \right> = \left< \tilde{q} \right> = \left< \tilde{L} \right> = \left< \tilde{\ell} \right> = 0$ and $\left< H \right> = m_h \slash \sqrt{ \lambda}$.  For these values of the coupling constants:
\begin{align}
&m_{\tilde{L}} = m_{\tilde{l}} = 10 \sqrt{2} \; \mathrm{GeV} & &m_H = .5 \sqrt{2} \; \mathrm{GeV}\nonumber \\
&m_{\tilde{Q}} = m_{\tilde{q}} = 15 \sqrt{2} \; \mathrm{GeV} & &A_L = 23 \; \mathrm{GeV} \nonumber \\
&  A_S = 31 \; \mathrm{GeV} & &\lambda = .006 \nonumber \\
& g_1= g_2 = .6 & &y = 1,
\end{align}
the global minimum is at the minimum mentioned above, while a local minimum occurs at $\left< \tilde{Q} \right> = \left< \tilde{q} \right> = 0$, $\left< \tilde{L} \right> =  5.025\; \mathrm{GeV}$, $ \left< \tilde{l} \right> = 5.136 \; \mathrm{GeV}$, and $\left< H \right> = 8.292 \; \mathrm{GeV} $.  Since the squarks do not acquire a VEV in this vacuum, baryon number is conserved and can be used to construct Q-balls.

In the false vacuum, the lightest squark $\tilde{q}^\prime$ has a mass $m_{\tilde{q}^\prime}$ of 7.80 GeV.  Because lepton number is not conserved, the sleptons mix with the Higgs boson; the lightest of these eigenstates is $h^\prime$ with a mass of 6.83 GeV.  We assume that the quark acquires a mass from the term $y \bar{q} H q$ in the Lagrangian; then it has mass $y\left< H \right> = 8.29 \; \mathrm{GeV}$, and so the lightest squark is stable against decay into a quark.

\begin{figure}
\includegraphics[scale=.7]{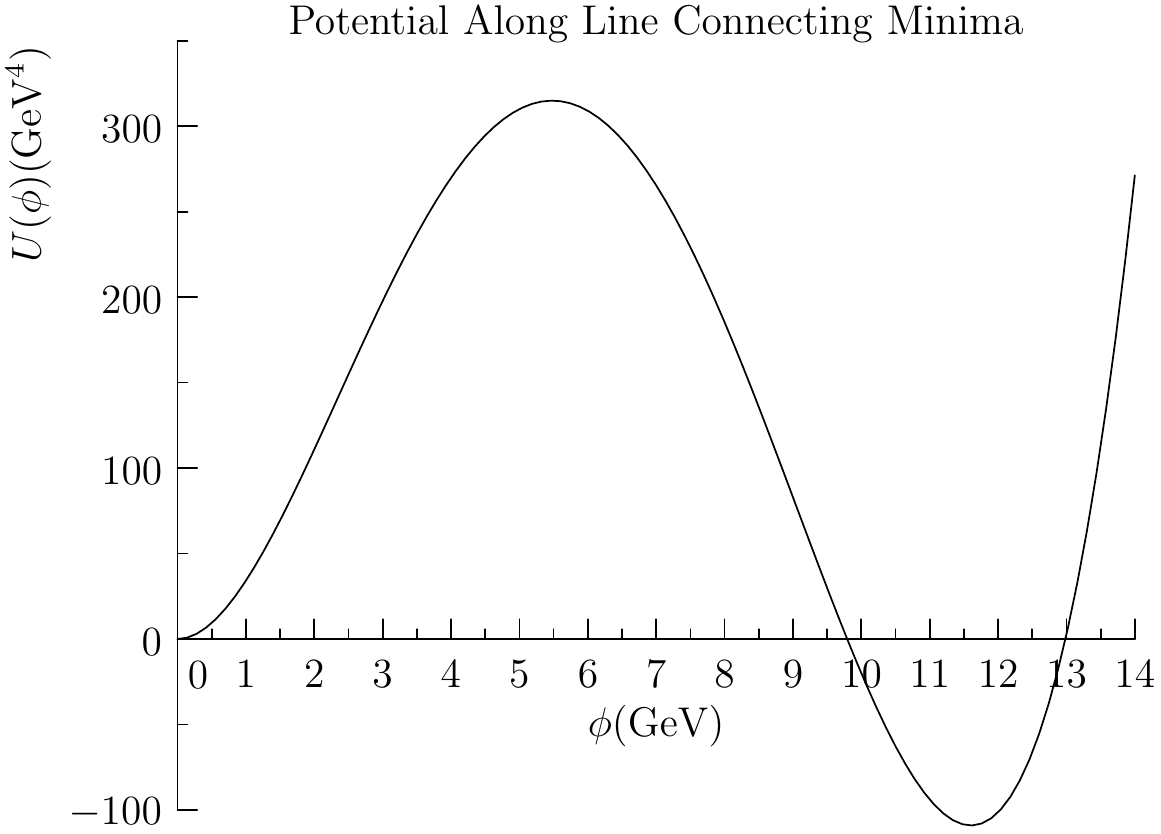}
\caption{The potential along the line connecting the false vacuum to the true vacuum.}
\label{fig:PotentialAlongLineConnectingMinima_MSSM}
\end{figure}

The potential along the line connecting the minima is
\begin{equation}
U(\phi) = .324 \phi^4 - 7.38 \; \mathrm{GeV} \; \phi^3 +41.2 \; \mathrm{GeV}^2 \; \phi^2,
\end{equation}
which is shown in Fig. \ref{fig:PotentialAlongLineConnectingMinima_MSSM}.  We notice that the barrier separating the local minimum from the global minimum is quite large, which dramatically suppresses tunneling through thermal fluctuations.  By expanding the potential in terms of the appropriate eigenstates in the false vacuum, one can show that Q-balls constructed of $\tilde{q}^\prime$ and $h^\prime$ fields do exist.  Thus we have all of the necessary ingredients for a solitosynthesis-induced phase transtion.  

While most of our analysis could be straightforwardly applied to this situation, the Bethe-Salpeter regime analysis cannot.  Firstly, because of the balance between the squark and quark masses, we do not have $m_{h^\prime} \ll m_{\tilde{q}^\prime}$ as required for the Bethe-Salpeter equation.  As yet, the Bethe-Salpeter equation for the exchange of massive particles is unsolved.  

Secondly, as was noted in the subsection ``Bethe-Salpeter Regime" of the section ``Properties of Q-balls in the False Vacuum", the field content of states described by the Bethe-Salpeter equation does not necessarily match the field content of Q-balls described by the thick wall approximation.  In our numerical example, we made the difference between the two small by choosing the global minimum such that $\sin(\theta) \approx 1$.  However, in this scenario, we must have $\sin(\theta) \ll 1$ because we must tunnel to a state near the global minimum, which has $ \left< \tilde{q}^\prime \right> = 0 $.  Therefore, the analysis should be modified to account for the fact that the Q-balls consist almost entirely of the $h^\prime$ field, with very little of the $\tilde{q}^\prime$ field.

However, even if such a phase transition could not have occured in the evolution of our universe, it is still important to study the regions of parameter space in the MSSM in which such a phase transition could occur.  Such a phase transition could destabilize a vacuum previously thought to be stable on cosmological timescales, leading to further constraints beyond those of~\cite{VacuumStability}.

\section{Conclusions}

We have considered the properties and growth of Q-balls in the false vacuum in each of four regimes, ranging from extremely small Q-balls to extremely large Q-balls, and we have demonstrated that phase transitions induced by solitosynthesis are indeed possible.  While we have used a toy model inspired by the MSSM, such phase transitions occur what any model with a similar potential and vacuum structure.

These phase transitions are of cosmological interest.  One unique aspect of these phase transitions is that the resulting vacuum carries a net charge afterwards; as mentioned in the introduction, this has been suggested as a possible baryogenesis mechanism in the Affleck-Dine mechanism.  However, more work must be done to establish that these phase transitions occur in the most promising MSSM potentials for this scenario.  Even if this cannot be established, the existence of such phase transitions may set new bounds on the allowed MSSM parameter space.

The author would like to thank Alex Kusenko for very helpful discussions.  This work was supported in part by DOE grant DE-FG03-91ER40662.

\appendix*
\section{Appendix: The Bethe-Salpeter Equation}

\begin{figure}
\includegraphics[scale=.8]{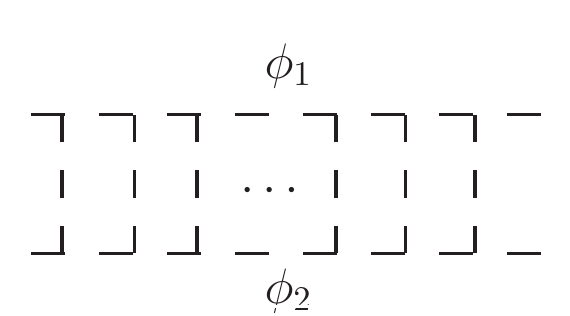}
\caption{These Feynman diagrams are considered in the ladder approximation of the Bethe-Salpeter equation.}\label{fig:Ladder}
\end{figure}

In this appendix, we will review the Bethe-Salpeter equation and derive the results given in the text for unequal masses in more detail.  The Bethe-Salpeter equation describes the relativistic bound states of a strongly interacting system, described by the ket $\left. | B \right>$.  The amplitude is:
\begin{equation}
\psi(x_1,x_2;P) = \left< 0 | T \phi_1(x_1) \phi_2(x_2) | B \right>,
\end{equation}
where $P$ is the four-monentum of the bound state.  One may simplify this equation by the ladder approximation, in which one considers only diagrams of the type shown in Fig. \ref{fig:Ladder}.  The Bethe-Salpeter equation has been solved in the Wick-Cutkosky model, in which the particles involved are two scalars exchanging massless quanta~\cite{Bethe Salpeter-same}.  In this model, the Bethe-Salpeter equation after a Wick rotation can be written as:
\begin{align}
\left[ \left( \dfrac{M}{2} + p \right)^2 + m_q^2 \right] &\left[ \left( \dfrac{M}{2} - p \right)^2 + m_q^2 \right] \psi(p) \nonumber \\
&= \dfrac{\tilde{A}^{2}}{16 \pi^4} \int d^4k \dfrac{\psi(k)}{(p-k)^2 + m_h^2},
\end{align}
where $M$ is the mass of the bound state, $m$ is the mass of the $\phi$ fields, and $A$ is the coupling between $\phi$ and the massless particle.  The momenta of the particles on the top and bottom are $P \slash 2 \pm p$.  The wavefunction can be found by solving a Friedholm differential equation~\cite{B-S Solution}; however, we need only the bound state masses, which are given by~\cite{Bethe Salpeter-same}:
\begin{equation} M_n = 2 m \left( 1 - \dfrac{\alpha^2}{8 n^2} \right), \label{eq:BSEqualMass} \end{equation}
if $ \alpha = A^{2} \slash 16 \pi m^2 < 1 $.

In the Wick-Cutkosky model, the particles at the top and bottom of the ladder are identical; however, this can be weakened.  Maintaining the ladder approximation, the Bethe-Salpeter equation for non-identical scalars interacting through massless exchange is:
\begin{align}
\left[ (m+\Delta)^2 + (p-\imath \eta_1 P)^2 \right] &\left[ (m-\Delta)^2 +(p + \imath \eta_2 P)^2 \right] \psi(p) \nonumber \\
&= \dfrac{A^{\prime}}{\pi^2} \int d^4k \dfrac{\psi(k)}{(p-k)^2},
\end{align}
where the masses of the particles on the top and bottom are $ m \pm \Delta $.  The coupling constant $ A^\prime $ is related to the coupling constants at the top and bottom interactions by $ A^\prime = g_{top} g_{bottom} \slash 16 \pi^2 $.  $ \eta_1 $ and $ \eta_2 $ come from transforming to a ``center of momentum" reference frame; these are:
\[ \eta_1 = \dfrac{m_{top}}{m_{top} + m_{bottom}} \qquad \eta_2 = \dfrac{m_{bottom}}{m_{top} + m_{bottom}}. \]
The energy-momentum four-vector of the bound state, $ P $, is given by $ (0,M) $ where $ M $ is the bound state mass.

As derived in~\cite{Bethe Salpeter}, this equation can be related to the Wick-Cutkosky model Bethe-Salpeter equation.  The result is that if $ A^{\prime} = F(M^2) $ in the case of equal masses, then for unequal masses:
\begin{equation} \dfrac{A^{\prime}}{1-\Delta^2\slash m^2} = F \left( \dfrac{M^2 - 4 \Delta^2}{1-\Delta^2\slash m^2} \right). \end{equation}

We will use this to derive the ground state for unequal masses.  First, we note that for equal masses $A^{\prime} = \tilde{A}^2 \slash 16 \pi^2$ since $g_{top} = g_{bottom} = \tilde{A}$.  Therefore, equation \eqref{eq:BSEqualMass} is:
\begin{equation} \dfrac{M}{2m} = 1 - \dfrac{A^{\prime \, 2} \pi^2}{8 m^4}. \end{equation}
Squaring this and keeping the lowest order terms using $A^\prime \ll m^2$ gives:
\begin{equation}
\dfrac{M^2}{4m^2} = 1 - \dfrac{A^{\prime \, 2} \pi^2}{4 m^4}
\end{equation}
which can be solved for:
\begin{equation}
A^{\prime} = \sqrt{\dfrac{4 m^4}{\pi^2} \left( 1 - \dfrac{M^2}{4m^2} \right)}.
\end{equation}

Then for unequal masses:
\begin{align}
\dfrac{A^{\prime} }{(1 - \Delta^2 \slash m^2 )} &= \sqrt{\dfrac{4 m^4}{\pi^2} \left( 1 - \dfrac{1}{4 m^2} \dfrac{M^2 - 4 \Delta^2}{1- \Delta^2 \slash m^2} \right)} \\
\dfrac{A^{\prime \, 2}}{(1 - \Delta^2 \slash m^2 )^2} &= \dfrac{4 m^4}{\pi^2} \left( 1 - \dfrac{1}{4 m^2} \dfrac{M^2 - 4 \Delta^2}{1- \Delta^2 \slash m^2} \right)
\end{align}
and solving for $ M^2 $ gives:
\begin{equation} M^2 = 4 \Delta^2 + 4 m^2 \left( 1 - \dfrac{\Delta^2}{m^2} \right) \left( 1 - \dfrac{A^{\prime \, 2} \pi^2}{4 m^4} \dfrac{1}{(1-\Delta^2 \slash m^2)^2} \right). \end{equation}
as in \eqref{UnequalMasses}.  Setting $\Delta = 0$ restores the equal mass case.  This remains valid as long as $A^\prime \ll m^2$.  If this fails, however, we are in a strong coupling regime in which the Bethe-Salpeter equation underestimates the binding energy; therefore Q-balls will be more likely to bind together than we find here.  Thus if we find a phase transition with this approximation, the phenomenon will still occur if the binding energies were calculated exactly.

\end{document}